\documentclass[
    aps,prl,
    twocolumn,
    reprint,
    superscriptaddress,
    nofootinbib,
    floatfix,
    amssymb,
    long bibliography
]{revtex4-1}

\usepackage[T1]{fontenc}
\usepackage[utf8]{inputenc} 
\usepackage[english]{babel}

\usepackage{graphicx}

\usepackage{mathtools}  
\usepackage{amsthm}
\usepackage{bm}         

\usepackage[normalem]{ulem}   
\usepackage[dvipsnames]{xcolor}

\usepackage[colorlinks=true,linkcolor=Blue,citecolor=Blue,urlcolor=Blue]{hyperref}
\usepackage[nameinlink,noabbrev]{cleveref} 

\renewcommand{\cite}{\citep}  
\newif\ifshownotes
\shownotestrue 

\ifshownotes
  \newcommand{\note}[3]{\textbf{\color{#2}[#1: #3]}}
\else
  \newcommand{\note}[3]{}
\fi

\begin{document}

\title{Active fluctuations induce buckling of living surfaces}
\author{Matteo Ciarchi}
\affiliation{Arnold Sommerfeld Center for Theoretical Physics and Center for NanoScience, Department of Physics, Ludwig-Maximilians-Universit\"at M\"unchen, Theresienstra\ss e 37, D-80333 M\"unchen, Germany}
\affiliation{Max Planck Institute for the Physics of Complex Systems, Nöthnitzer Straße 38, D-01187 Dresden, Germany}
\author{Andriy Goychuk}
\affiliation{Arnold Sommerfeld Center for Theoretical Physics and Center for NanoScience, Department of Physics, Ludwig-Maximilians-Universit\"at M\"unchen, Theresienstra\ss e 37, D-80333 M\"unchen, Germany}
\affiliation{Department of Systems Immunology, Helmholtz Centre for Infection Research, 38124 Braunschweig, Germany}
\affiliation{Lower Saxony Center for Artificial Intelligence and Causal Methods in Medicine (CAIMed), Hannover, Germany}
\affiliation{Institute for Biochemistry, Biotechnology and Bioinformatics, Technische Universität Braunschweig, Braunschweig, Germany}
\author{Erwin Frey}
\email{frey@lmu.de}
\affiliation{Arnold Sommerfeld Center for Theoretical Physics and Center for NanoScience, Department of Physics, Ludwig-Maximilians-Universit\"at M\"unchen, Theresienstra\ss e 37, D-80333 M\"unchen, Germany}
\affiliation{Max Planck School Matter to Life, Hofgartenstraße 8, D-80539 Munich, Germany}

\date{\today} 

\begin{abstract}
Active tissues exhibit tension fluctuations that are correlated in space and time. 
We study a minimal overdamped surface model in which such fluctuations enter as a zero-mean, multiplicative modulation of the local surface tension.
Although the deterministic elastic dynamics (tension plus bending) stabilizes the flat state for all nonzero wave numbers, we find that sufficiently persistent active fluctuations generate positive ensemble growth rates for a finite band of Fourier modes, leading to stochastic buckling with wavelength selection.
A non-Markovian theory based on the Novikov--Furutsu theorem captures the instability threshold and unstable band observed in simulations.
\end{abstract}


\maketitle

Non-equilibrium fluctuations are a central feature of active and biological matter, where chemical energy is continuously converted into mechanical work and stress.
For example, epithelial tissues show collective fluctuations in cell density and waves~\cite{Article::Deforet2014, Article::Zehnder2015, Article::Zehnder2015b}.
These tissue dynamics are driven by cell migration~\cite{Peyret_2019} and feedback mechanisms due to chemical signaling~\cite{Article::heer2017tension, Article::Boocock2020}, coupled with stress propagation and non-Gaussian stress fluctuations~\cite{Article::Trepat2009, Article::Tambe2011, Article::Serra-Picamal2012}.
In an active-matter setting, actin filaments propelled by surface-bound motors in gliding assays exhibit strongly non-thermal bending fluctuations: the filament-curvature distribution develops an exponential tail, reflecting intermittent, jump-like motor forcing rather than a simple effective-temperature renormalization~\cite{Weber2015}.

In general, nonequilibrium fluctuations in the form of \emph{multiplicative noise} can qualitatively change or even generate spatio-temporal structures in nonequilibrium systems through distinct mechanisms. 
In ecological and evolutionary games, intrinsic demographic noise together with dispersal can organize coherent spatiotemporal patterns (e.g., spiral waves) with a selected wavelength~\cite{Reichenbach:2007a, Reichenbach:2007b}.
Demographic fluctuations can also sustain pronounced spatial correlations even when deterministic dynamics is linearly stable, by continually exciting the slowest-decaying modes~\cite{Article::biancalani2017giant, Article::butler2009robust}.
In nonlinear media, multiplicative noise can promote spatial patterning, with emergent length scales controlled by the noise statistics~\cite{Sagues:2007}.

Here, we identify a distinct route to noise-induced pattern formation. 
We consider a minimal overdamped elastic surface with strictly positive mean tension, for which the deterministic dynamics relaxes \emph{all} finite-wavelength perturbations. 
We show that a \emph{zero-mean}, spatiotemporally correlated \emph{multiplicative} modulation of the tension can nevertheless generate a genuine \emph{ensemble-level} instability: a finite band of Fourier modes acquires positive growth rates, leading to \emph{stochastic buckling} and \emph{robust wavelength selection}. 
Importantly, this instability is absent in every deterministic realization and is not ``imprinted'' in the noise correlations. 
Instead, it arises from the interplay of colored multiplicative fluctuations with elastic relaxation, which induces an effective memory feedback in the averaged dynamics. 
We support these predictions by stochastic simulations and by an analytical theory that yields the dispersion relation and instability threshold.


\paragraph*{Model.}
To make this mechanism explicit, we consider the simplest setting in which activity enters solely through a fluctuating tension.
We study an overdamped height field ${h(\boldsymbol{x},t)}$ in Monge gauge (small slopes), whose elastic relaxation is modulated by nonequilibrium fluctuations of the local surface tension. 
We model these fluctuations as arising from areal-density variations ${\delta\rho(\boldsymbol{x},t)}$ that shift the tension according to ${\delta\sigma(\boldsymbol{x},t)=-\alpha\,\delta\rho(\boldsymbol{x},t)}$, so that activity enters as a \emph{multiplicative}, spatially and temporally correlated modulation of the elastic operator. 
Neglecting additive thermal noise to isolate this effect, the dynamics reads
\begin{equation}
\xi\,\partial_t h
= \boldsymbol{\nabla}\cdot\Big[(\bar{\sigma}+\delta\sigma)\,\boldsymbol{\nabla} h\Big]
- \kappa\,\boldsymbol{\nabla}^{4} h,
\label{eq:phys_eom}
\end{equation}
with bending rigidity ${\kappa}$, mean tension ${\bar{\sigma}>0}$, and viscous drag coefficient ${\xi}$.
In contrast to prior work~\cite{ramaswamy2000nonequilibrium, Gov2004}, all forces are internal to the surface and there are no kicks perpendicular to the surface.

We measure lengths by the elastocapillary scale ${\ell_p=\sqrt{\kappa/\bar{\sigma}}}$, times in units of ${\tau_r=\xi\kappa/\bar{\sigma}^2}$, and define the dimensionless tension fluctuation ${\sigma\equiv\delta\sigma/\bar{\sigma}}$.
In these units Eq.~\eqref{eq:phys_eom} can be written as gradient flow ${\partial_t h=-\delta\mathcal{F}[h;\sigma]/\delta h}$ with a \emph{time-dependent} quadratic free-energy functional~\cite{helfrich1, canham1970minimum} 
\begin{equation}
\mathcal{F}[h;\sigma]=\int d\boldsymbol{x}\,
\Big[\tfrac12(1+\sigma)(\boldsymbol{\nabla}h)^2+\tfrac12(\boldsymbol{\nabla}^2 h)^2\Big].
\end{equation}
In the absence of activity (${ \sigma=0 }$) the flat state is \emph{deterministically linearly stable}:
each Fourier mode decays with rate ${\bar{\sigma}q^2+\kappa q^4}$, while the mode ${q=0}$ is neutral by height-shift symmetry. 
The central question is whether a \emph{zero-mean}, correlated multiplicative modulation of tension can nevertheless destabilize long wavelengths and produce wavelength selection, despite deterministic stability for all $q\neq 0$.

\paragraph*{Active tension fluctuations.}
Motivated by the coherent, tissue-scale stress fluctuations observed in epithelia~\cite{Article::Trepat2009,Article::Tambe2011,Article::Serra-Picamal2012,Article::Deforet2014,Article::Zehnder2015,Article::Zehnder2015b,Article::Boocock2020}, we model the active tension as a zero-mean, stationary field with translationally invariant covariance ${\langle \sigma(\boldsymbol{x},t)\,\sigma(\boldsymbol{x}',t')\rangle \equiv \Gamma(\boldsymbol{x}-\boldsymbol{x}',t-t')}$ characterized by a correlation length ${\lambda}$ and persistence time ${\tau}$. 
For analytical and numerical convenience we take ${\sigma}$ to be Gaussian, noting that tissue stresses are in reality markedly non-Gaussian~\cite{Article::Trepat2009}; our conclusions rely only on finite variance and correlation scales, thereby uncoupling the fluctuations from the dissipation~\cite{Betz_2009, Turlier_2016, Gnesotto_2018} in Eq.~\eqref{eq:phys_eom}, not on Gaussianity.
Specifically, we prescribe the spectrum ${\Gamma_{qk}(t-t') \equiv \langle \sigma_q(t)\,\sigma_k(t')\rangle}$ for the Fourier modes $\sigma_q(t)$: 
\begin{equation}
    \Gamma_{qk}(t-t')
= \frac{\epsilon\,V\,\exp\!\left[-(1+\lambda^{2}q^{2})\,|t-t'|/\tau\right]}
{\tau\,(1+\lambda^{2}q^{2})}\,\delta_{q,-k}
\, ,
\label{eq:noisecorrfourier}
\end{equation}
where ${V}$ is the system volume and ${\epsilon}$ sets the overall noise strength.

\paragraph*{Heuristic: transient negative tension.}
For long but finite wavelengths $q>0$ and $\sigma(t)$ uniform, linear mode analysis shows that a mode amplifies over a window $\Delta t$ if ${\langle\sigma\rangle_{\Delta t}\!=\!\Delta t^{-1}\!\int_t^{t+\Delta t}\! dt' \,\sigma(t') < - (1+q^2)}$; this can be seen by making the mean-field approximation $\sigma \approx \langle \sigma \rangle_{\Delta t}$ in Eq.~\eqref{eq:phys_eom}. 
During such intervals the unstable band is ${0<q<\sqrt{|1+\langle\sigma\rangle_{\Delta t}|}}$  (with ${q=0}$ neutral). 
This qualitative argument shows that temporally correlated tension fluctuations can generate a transient instability.
A short-time cumulant expansion of the height-field dynamics supports this picture, as it predicts a dynamical instability beyond a critical wavelength $q^*$, with modes in the unstable band growing at early times~\cite{supp}.
Whether this short-time growth persists and culminates in an ensemble-level buckling transition with wavelength selection cannot be decided at this level. 
We therefore turn to stochastic simulations of Eq.~\eqref{eq:phys_eom} driven by the colored tension noise defined in Eq.~\eqref{eq:noisecorrfourier}.
Emergent structure is characterized by the equal-time height-correlation function ${C(x) \equiv \langle h(x) \, h(0)\rangle}$ and, equivalently, the structure factor ${S(q) \equiv \langle |h_q|^2\rangle}$.

\paragraph*{Stochastic simulations.}
We integrated Eq.~\eqref{eq:phys_eom} with colored tension fluctuations generated in Fourier space via an auxiliary Ornstein--Uhlenbeck (OU) process for each mode $\sigma_q(t)$ chosen such that $\Gamma_{qk} (t)$ matches Eq.~\eqref{eq:noisecorrfourier}; see~\cite{supp} for details.
Time stepping was performed with a semi-implicit scheme implemented in XMDS2~\cite{xmds2}. 
Unless stated otherwise, we set $\tau=0.2$ and $\lambda=1.0$. 
Simulations were run up to ${T = 20}$ with time step ${dt = 2.5\times10^{-4}}$ on a spatial grid of 2000 points with $dx=0.1$; we recorded the height $h$ at $500$ uniformly spaced time points, and averaged over $5\times10^4$ realizations.
To prevent runaway growth once unstable modes are excited, we added a local saturating nonlinearity $-h^3$ (equivalently, a local potential $V(h)\propto h^4$). 
Physically, this term may be viewed as an effective coupling of the surface to a substrate/cortex that penalizes large deformations and regularizes the nonlinear regime.

\begin{figure}[t]
\includegraphics[width=\columnwidth]{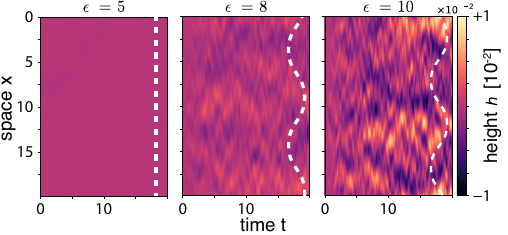}
\caption{%
\textbf{Space--time kymographs} of the height field $h(x,t)$ from simulations of Eq.~\eqref{eq:phys_eom} with colored tension noise ($\lambda=1$, $\tau=0.2$), shown for noise strength $\epsilon=5,8,10$ (left to right) on a common color scale.
White dashed curves serve as a guide to the eye, highlighting the late-time spatial modulation at larger noise strength.
}
\label{fig:heatmap_dyn_eps}
\end{figure}

Figure~\ref{fig:heatmap_dyn_eps} shows space--time kymographs of the height field $h(x,t)$ for increasing noise strength $\epsilon$.
For weak noise ($\epsilon=5$), fluctuations remain small and do not organize into a persistent spatial modulation. 
Above a threshold (here $\epsilon\gtrsim 8$), coherent undulations build up at late times and persist, indicating noise-induced stochastic buckling (dashed curves are guides to the eye).

\begin{figure}[ht]
\includegraphics[width=\columnwidth]{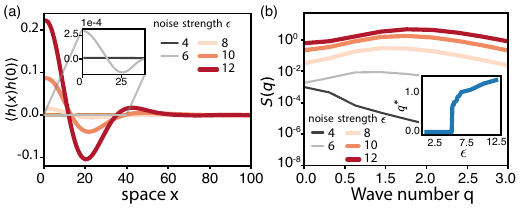}
\caption{
(a) Equal-time height correlation $C(x)=\langle h(x)h(0)\rangle$ at $t=11.6$ from simulations of Eq.~\eqref{eq:phys_eom} with colored tension noise [Eq.~\eqref{eq:noisecorrfourier}] of correlation length $\lambda=1$ and persistence time $\tau=0.2$.
Curves correspond to noise intensities $\epsilon=4,6,8,10,12$ (legend) and are averaged over $5\times10^{4}$ independent realizations.
Inset: same data with expanded vertical scale (note $10^{-4}$), highlighting the weak-noise regime. 
(b) Structure factor $S(q)=\langle |h_q|^2 \rangle$ obtained from the Fourier transform of $C(x)$ in (a), averaged over the last 100 saved time points of the dynamics. 
Inset: dominant wave number $q^*$ as a function of noise intensity. 
}
\label{fig:Cht}
\end{figure}

Consistently, the equal-time height-correlation function $C(x)$ (Fig.~\ref{fig:Cht}(a)) crosses over from monotone decay below threshold to damped oscillations above threshold, signaling wavelength selection. 
Notably, the selected wavelength exceeds the tension-noise correlation length $\lambda$ by more than an order of magnitude.

Taken together, the simulations reveal an activity-driven transition from a featureless, relaxing interface to a stochastically buckled state with a well-defined length scale.
We quantify wavelength selection by extracting the dominant wave number $q^\star$ from the maximum of the structure factor $S(q)$ [Fig.~\ref{fig:Cht}(b)] (or equivalently from the oscillation period of $C(x)$). 
We locate the onset of pattern formation by the noise strength $\epsilon$ at which $S(q,t)$ indicates such a maximum. 

Because the added nonlinearity serves only to saturate the amplitude, we expect both $q^\star$ and the threshold to be set by the \emph{linear} stochastic dynamics. 
We thus turn to an analytical theory for the mode-resolved \emph{ensemble growth rates} generated by colored multiplicative tension fluctuations.

\paragraph*{Theory: non-Markovian mean-field dynamics.}
Having observed a steady-state definite length scale in simulations, we now develop an analytical description of how colored multiplicative tension fluctuations generate a long-time instability and wavelength selection.
We consider the noise-averaged evolution of a single Fourier mode $\langle h_q(t)\rangle$. 
Since translational symmetry and unbiased initial conditions enforce $\langle h_q\rangle=0$ in an unseeded ensemble, we interpret $\langle h_q(t)\rangle$ as a \emph{linear-response} observable: the averaged evolution of an infinitesimally seeded mode (fixed phase) under the same stochastic driving. 
A positive growth rate of this response identifies the modes that eventually dominate $S(q)$ once nonlinear saturation limits the amplitude.

In Fourier space the linear stochastic dynamics can be written as ${\partial_t  h(t)=[J+\Lambda(t)] \, h(t)}$, with deterministic $J$ and a stochastic term $\Lambda(t)$ linear in $\sigma$ (see~\cite{supp}). 
The formal solution $h(t)=U(t,t_0) \, h(t_0)$ involves the propagator $U(t,t_0)=\mathcal{T}\exp \big[ \int_{t_0}^{t}dt'\,(J+\Lambda(t'))\big]$ where $\mathcal{T}$ denotes time ordering. 
Averaging Eq.~\eqref{eq:phys_eom} generates mixed moments such as $\langle \sigma\,h\rangle$. 
For Gaussian colored tension noise, the Novikov--Furutsu theorem~\cite{novikov, furutsu} expresses these terms exactly in terms of the response functional $\langle \delta h/\delta\sigma\rangle$ (see~\cite{supp}). 
Formally, $\delta h/\delta\sigma$ depends on the (time-ordered) propagator $U(t,s)$; an exact closure would thus require averages of time-ordered exponentials. 
We therefore expand to first order in the noise intensity $\epsilon$, evaluating the response with the \emph{unperturbed} deterministic propagator (equivalently, setting $\Lambda=0$ in the
response; see~\cite{supp}).

To $\mathcal{O}(\epsilon)$, each mode obeys a linear but non-Markovian equation with a memory kernel $K_q$,
\begin{equation}
\partial_t \langle h_q(t)\rangle
= J_q\,\langle h_q(t)\rangle
+\frac{\epsilon q^2}{\tau}\int_{t_0}^{t}\!dt'\,K_q(t-t')\,\langle h_q(t')\rangle,
\label{eq:effective_mean_field}
\end{equation}
where $J_q=-(q^2+q^4)$ is the deterministic relaxation rate. 
Although derived for small $\epsilon$, Eq.~\eqref{eq:effective_mean_field} retains the emergent memory kernel and thus captures the cumulative long-time influence of tension fluctuations on $\langle h_q\rangle$.
The kernel $K_q$ has a transparent mode-coupling interpretation: tension fluctuations at wave number $n$ scatter height fluctuations into mode $q$, with weights set by the spatiotemporal noise spectrum. For the Ornstein--Uhlenbeck spectrum in Eq.~\eqref{eq:noisecorrfourier} one obtains
\begin{equation}
K_q(t)=\int\!\frac{dn}{2\pi}\,
\frac{(n+q)^2}{1{+}\lambda^2 n^2}\,
\exp\!\Big[\frac{\tau J_{n+q}-(1{+}\lambda^2 n^2)}{\tau}\,t\Big].
\end{equation}
Equation~\eqref{eq:effective_mean_field} makes the mechanism explicit: colored multiplicative noise induces an effective feedback with memory that can overcome deterministic relaxation and destabilize a finite band of wave numbers, enabling wavelength selection.

\begin{figure}[t]
    \centering
    \includegraphics{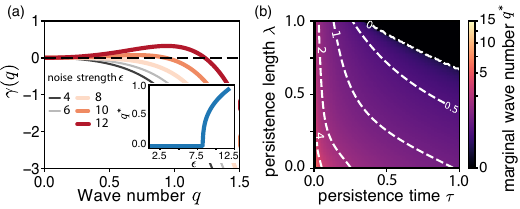}
\caption{%
Properties of unstable modes predicted by the non-Markovian mean-field theory.
(a) Mode growth rate $\gamma(q)$ (shown as $\partial_t\langle h_q\rangle|_{t=0}$ for a unit seed) as a function of wave number $q$ for the indicated noise strengths $\epsilon$ at fixed ${\tau=0.2}$ and ${\lambda=1}$. 
The dashed line marks marginal stability, ${\gamma=0}$; a finite
band with ${\gamma(q)>0}$ appears for ${\epsilon\gtrsim 8}$. 
Inset: dominant wave number as a function of noise strength $\epsilon$.
(b) Marginal wave number $q^{\star}$ (color code) at onset
(${\max_q\gamma(q)=0}$) as a function of $\tau$ and $\lambda$ for ${\epsilon=10}$.
White dashed lines are contours of constant $q^{\star}$ (values displayed on dashed line).}
\label{fig:disp}
\end{figure}

\paragraph*{Dispersion relation and comparison to simulations.---}
By applying the Laplace transform $\langle h_q(t) \rangle \rightarrow \langle \tilde{h}_q(s) \rangle =\int_0^\infty dt e^{-st}\langle h_q(t)\rangle$ to Eq.~\eqref{eq:effective_mean_field} and expanding the right-hand side in the Laplace parameter $s$ to first order, we obtain the dispersion relation $\gamma(q)$ as the simple pole in $s$ of the resulting solution for $\langle \tilde{h}_q(s) \rangle=\frac{A(q,\lambda,\tau)}{s-\gamma(q)}$.
Consequently, at long times $\langle h_q(t)\rangle\sim e^{\gamma(q)t}$. 
A buckling transition occurs when $\max_q \gamma(q)$ crosses zero; the selected wave number $q^\star$ is the maximizer of $\gamma(q)$ (and coincides with the peak of $S(q)$ in the saturated state).
Figure~\ref{fig:disp} summarizes the resulting predictions: (i) $\gamma(q)$ develops a \emph{finite} unstable band, (ii) the threshold $\epsilon_c$ for which $\max_q\gamma(q)=0$ is consistent with the simulations, and (iii) the selected wave number $q^\star$ depends systematically on the noise correlation length $\lambda$ and persistence time $\tau$. 
Increasing $\lambda$ so that tension fluctuations are spatially smoother and/or $\tau$ so that tension fluctuations are slower suppresses the instability.

These trends are borne out in simulations.
Above threshold, we observe the emergence of a robust spatial modulation in $h(x,t)$ and, correspondingly, damped oscillations in the
equal-time correlation $C(x)$ [Fig.~\ref{fig:Cht}(a)] together with a pronounced peak in $S(q)$
[Fig.~\ref{fig:Cht}(b)].
As $\epsilon$ is increased above onset, the peak shifts to larger $q$ (shorter wavelength), consistent with the growth-rate maximum $q^{*}$ extracted from $\gamma(q)$.

The theory is controlled for small $\epsilon$ and is expected to lose quantitative accuracy deep in the nonlinear regime. 
For large noise strengths many modes are simultaneously excited, and higher-order contributions in $\epsilon$ generate additional mode couplings beyond the leading-order closure in Eq.~\eqref{eq:effective_mean_field}.
Nevertheless, simulations show that wavelength selection persists at higher $\epsilon$, indicating that the non-Markovian feedback mechanism identified here captures the essential physics of the instability.

\paragraph*{Markovian approximation.}
To derive an analytical expression for the dispersion relation, we will now explore the asymptotic limit $t \gg \tau$ in which Eq.~\eqref{eq:effective_mean_field} effectively becomes memoryless.
This limit allows approximating the exponential in the memory kernel $K_q(t)$ as a half-sided delta-distribution $\delta_{+}(t)$, leading to
\begin{equation}
K_q(t) \approx \int\!\frac{dn}{2\pi}\,
\frac{(n+q)^2}{1+\lambda^2 n^2}\,
\frac{\tau}{1+\lambda^2 n^2 - \tau J_{n+q}} \, \delta_{+}(t)
.
\label{eq:markovian_approximation}
\end{equation}
Evaluating the integral to lowest order in $\tau$ and substituting in Eq.~\eqref{eq:effective_mean_field} leads to
\begin{equation}
\partial_t \langle h_q(t)\rangle
= \left[J_q
+ \epsilon \, \frac{q^2 + \lambda^2 q^4}{4 \lambda^3} \right] \,\langle h_q(t)\rangle .
\end{equation}
Thus, in the Markovian limit $\tau \to 0$ our analysis based on the Novikov-Furutsu theorem converges to the cumulant expansion~\cite{supp}.
Conversely, evaluating Eq.~\eqref{eq:markovian_approximation} to lowest order in $\tau$ for noise lacking spatial correlations, $\lambda = 0$, and substituting in Eq.~\eqref{eq:effective_mean_field},
\begin{equation}
\partial_t \langle h_q(t)\rangle
= \left[J_q + \frac{\epsilon q^2}{2\sqrt{2} \tau^{3/4}} \right] \langle h_q(t)\rangle,
\end{equation}
reveals that the patterns are not imprinted by the spatial structure of the noise.

\paragraph*{Summary and conclusions.---} 
In summary, we have shown that spatiotemporally correlated, \emph{multiplicative coloured noise} can qualitatively reshape the long-time dynamics of a spatially extended system. 
Even when the deterministic dynamics is strictly relaxing---here, an overdamped surface with
$\bar{\sigma}>0$ and bending rigidity---colored tension fluctuations can generate an \emph{ensemble-level} instability with a finite unstable band, leading to stochastic buckling and wavelength selection.

Our findings illustrate a distinct mechanism towards noise-induced pattern formation, which differs from prior literature as follows.
In contrast to noise-sustained Turing-like patterns, where the deterministic operator already possesses a least-damped finite-$q$ band and a stochastic forcing continually excites these near-marginal modes~\cite{Article::butler2009robust, Article::biancalani2017giant}, we find wavelength selection in a system whose passive response is monotone in wave number $q$.
Zero-mean colored \emph{multiplicative} tension fluctuations induce an effective memory feedback in the averaged dynamics, destabilizing a finite band of modes and yielding patterns.
In contrast to previous work~\cite{Article::adamer2020coloured, Article::sanz2001turing, Goychuk_2023, Goychuk_2024}, these patterns are not imprinted by the noise correlations \emph{per se}.

Beyond active surfaces, the same mechanism should arise whenever a stable linear dynamics is modulated by correlated parameter fluctuations, including evolutionary dynamics on fluctuating fitness landscapes~\cite{Loffler_2020} and other high-dimensional biological and ecological systems~\cite{Liu,Spagnolo,Coomer}.
More generally, our non-Markovian framework provides a practical route to predict noise-driven instability thresholds and selected length scales directly from the spatiotemporal statistics of the driving.

\begin{acknowledgments}
We thank Mehran Kardar and Arup K. Chakraborty for stimulating discussions and for pointing us towards work on evolution on fluctuating seascapes. 
E.F. was supported by the European Research Council (ERC) under the European Union’s Horizon Europe programme (CellGeom, Grant No. 101097810), the Templeton Foundation, and the Deutsche Forschungsgemeinschaft (DFG, German Research Foundation) under Germany’s Excellence Strategy—EXC-2094—390783311 (Excellence Cluster ORIGINS). 
A.G. was supported by the Ministry of Science and Culture of Lower Saxony via zukunft.niedersachsen (Volkswagen Foundation) within CAIMed—Lower Saxony Center for Artificial Intelligence and Causal Methods in Medicine.
Part of this work was performed at the Massachusetts Institute of Technology, where A.G. was supported by an EMBO Postdoctoral Fellowship (ALTF 259-2022).
\end{acknowledgments}

\bibliography{bibl}
\end{document}


\title{\text{\small Supplementary Information for} \\
Active fluctuations induce buckling of living surfaces}
\author{Matteo Ciarchi}
\affiliation{Arnold Sommerfeld Center for Theoretical Physics and Center for NanoScience, Department of Physics, Ludwig-Maximilians-Universit\"at M\"unchen, Theresienstra\ss e 37, D-80333 M\"unchen, Germany}
\affiliation{Max Planck Institute for the Physics of Complex Systems, Nöthnitzer Straße 38, D-01187 Dresden, Germany}
\author{Andriy Goychuk}
\affiliation{Arnold Sommerfeld Center for Theoretical Physics and Center for NanoScience, Department of Physics, Ludwig-Maximilians-Universit\"at M\"unchen, Theresienstra\ss e 37, D-80333 M\"unchen, Germany}
\affiliation{Department of Systems Immunology, Helmholtz Centre for Infection Research, 38124 Braunschweig, Germany}
\affiliation{Lower Saxony Center for Artificial Intelligence and Causal Methods in Medicine (CAIMed), Hannover, Germany}
\affiliation{Institute for Biochemistry, Biotechnology and Bioinformatics, Technische Universität Braunschweig, Braunschweig, Germany}
\author{Erwin Frey}
\email{frey@lmu.de}
\affiliation{Arnold Sommerfeld Center for Theoretical Physics and Center for NanoScience, Department of Physics, Ludwig-Maximilians-Universit\"at M\"unchen, Theresienstra\ss e 37, D-80333 M\"unchen, Germany}
\affiliation{Max Planck School Matter to Life, Hofgartenstraße 8, D-80539 Munich, Germany}

\maketitle

\clearpage\newpage

\subsection*{Formal solution and short-time cumulant/Magnus analysis}


After Fourier transforming the height field, and collecting modes into $\vec h(t)=\{h_{\mathbf q}(t)\}$, we can write Eq.~\eqref{main-eq:phys_eom} as
\begin{equation}
\partial_t \vec h(t)=\big[\mat J+\mat \Lambda(t)\big]\vec h(t),
\label{eq:app-operator}
\end{equation}
where we have defined
\begin{equation}
J_{\mathbf q\mathbf k}= -\big(|\mathbf q|^2+|\mathbf q|^4\big)\,\delta_{\mathbf q\mathbf k},\quad
\Lambda_{\mathbf q\mathbf k}(t)= -\frac{\mathbf q\!\cdot\!\mathbf k}{V}\,\sigma_{\mathbf q-\mathbf k}(t),
\label{eq:app-stochastic_matrix}
\end{equation}
and where $V$ is the system volume (in 1D, $V=L$). 
Here $\mat J$ is the deterministic linear operator; $\mat\Lambda(t)$ is the (multiplicative) stochastic “vertex” linear in the tension fluctuations.

Let $\mat U(t,t_0)$ solve ${\partial_t \mat U=(\mat J+\mat\Lambda)\mat U}$ with ${\mat U(t_0,t_0)=\mat I}$.
Iterating Duhamel’s formula gives the Dyson series, compactly  \cite{vanKampenBook}:
\begin{equation}
\vec h(t)=\underbrace{\Big\lceil\exp\!\Big\{\int_{t_0}^{t}\!\!dt'\,[\mat J+\mat\Lambda(t')]\Big\}\Big\rceil}_{\displaystyle \mat U(t,t_0)}\,\vec h(t_0),
\label{eq:app-formal}
\end{equation}
where time ordering $\lceil\cdots\rceil \equiv \mathcal{T}\cdots$ is needed because matrices at different times do not commute.

Factoring out the deterministic relaxation via ${\vec u(t):=e^{-\mat J(t-t_0)}\vec h(t)}$, one has
\begin{align}
\partial_t\vec u(t)&=\mat\Lambda_I(t)\,\vec u(t),\\
\mat\Lambda_I(t)&:=e^{-\mat J(t-t_0)}\mat\Lambda(t)\,e^{\mat J(t-t_0)}.
\end{align}
With $\omega_{\mathbf q}:=|\mathbf q|^2+|\mathbf q|^4$, the interaction-picture matrix reads
\begin{equation}
\big[\Lambda_I(t)\big]_{\mathbf q\mathbf k}
= -\frac{\mathbf q\!\cdot\!\mathbf k}{V}\,\sigma_{\mathbf q-\mathbf k}(t)\,
e^{\,(\omega_{\mathbf q}-\omega_{\mathbf k})(t-t_0)}.
\label{eq:app-LambdaI}
\end{equation}
Intuition: $\mat J$ sets the bare decay $e^{-\omega_{\mathbf q}t}$; $\mat\Lambda_I$ injects mode coupling weighted by the instantaneous stress and by the frequency mismatch.

We seek the mean response 
\[
\vec m(t):=\langle\vec h(t)\rangle=e^{\mat J(t-t_0)}\langle\vec u(t)\rangle
\]
for zero-mean, Gaussian $\sigma$ with correlator [Eq.~(2) in the main text]
\begin{equation}
\langle \sigma_{\mathbf p}(t)\sigma_{\mathbf p'}(t')\rangle
=\frac{\varepsilon}{\tau\,[1+\lambda^2 p^2]}\,
e^{-\,[1+\lambda^2 p^2]\lvert t-t'\rvert/\tau}\;V\,\delta_{\mathbf p,-\mathbf p'}.
\label{eq:app-corr}
\end{equation}
Because $\langle \sigma\rangle=0$, the first nonzero contribution in a cumulant (van Kampen) or Magnus expansion is \emph{second order} in $\mat\Lambda_I$:
\begin{equation}
\big\langle \vec u(t)\big\rangle
\simeq \exp\!\Biggl[\int_{t_0}^{t}\!\!dt_1\!\int_{t_0}^{t_1}\!\!dt_2\;
\big\langle \mat\Lambda_I(t_1)\mat\Lambda_I(t_2) \big\rangle \Biggr]\vec u(t_0),
\label{eq:app-cumulant}
\end{equation}
leading to an effective generator $\mat J+\mat\Sigma$ for $\langle \vec h(t)\rangle$:
\begin{align}
\partial_t \langle \vec h(t)\rangle
&=\Big[\mat J+\mat\Sigma\Big]\langle \vec h(t)\rangle,\\
\mat\Sigma&=\int_{0}^{t-t_0}\!d u\;\big\langle \mat\Lambda(t)\, e^{\mat{J}u}\mat\Lambda(t-u)e^{-\mat{J}u}\big\rangle,
\label{eq:app-Sigma-def}
\end{align}
valid for $\tau\ll t\ll \tau_r$ (noise correlation time $\tau$ much shorter than the deterministic relaxation time $\tau_r$).

We now proceed by making the following simplifications.
First, noting that the short time window $t-t_0 \ll \tau_r$ implies that there is no time for $\mat{J}$ to act, we approximate $e^{\pm\mat{J}u} \approx \mat{I}$ in Eq.~\eqref{eq:app-Sigma-def}.
Then, we effectively take $t - t_0 \to \infty$ because $\tau$ is so small that the noise decorrelates quickly.

With these simplifications, one has
\begin{equation}
\mat\Sigma_{\mathbf{q}\mathbf{k}} = \int_{0}^{\infty}\!d u\; \sum_{\mathbf{p}} \big\langle \mat\Lambda_{\mathbf{q}\mathbf{p}}(t)\,\mat\Lambda_{\mathbf{p}\mathbf{k}}(t-u)\big\rangle.
\end{equation}
Substituting the stochastic 'vertex', which is defined in Eq.~\eqref{eq:app-stochastic_matrix}, leads to:
\begin{equation}
\mat\Sigma_{\mathbf{q}\mathbf{k}} = \int_{0}^{\infty}\!d u\; \sum_{\mathbf{p}} \frac{\mathbf q\!\cdot\!\mathbf p}{V}\, \frac{\mathbf p\!\cdot\!\mathbf k}{V}\, \big\langle \sigma_{\mathbf q-\mathbf p}(t)\,\sigma_{\mathbf p-\mathbf k}(t-u)\big\rangle.
\end{equation}
Substituting the correlator of the noise [Eq.~\eqref{eq:app-corr}], one has:
\begin{equation}
\mat\Sigma_{\mathbf{q}\mathbf{k}} = \int_{0}^{\infty}\!d u\; \frac{1}{V} \sum_{\mathbf{p}} \frac{\epsilon (\mathbf q\!\cdot\!\mathbf p)^2}{\tau(\lambda^2 |\mathbf{q}-\mathbf{p}|^2 +1)} \, 
    e^{- (\lambda^2 |\mathbf{q}-\mathbf{p}|^2+1) u / \tau} \, \delta_{\mathbf{q}\mathbf{k}}.
\end{equation}
Finally, after time integration, one arrives at
\begin{equation}
\mat\Sigma_{\mathbf{q}\mathbf{k}} = \frac{1}{V} \sum_{\mathbf{p}} \frac{\epsilon (\mathbf q\!\cdot\!\mathbf p)^2}{(\lambda^2 |\mathbf{q} -\mathbf{p}|^2 +1)^2} \, \delta_{\mathbf{q}\mathbf{k}}.
\end{equation}
For large system sizes, one can approximate the sum as an integral in momentum space which, after a change of variables, gives
\begin{equation}
\mat\Sigma_{\mathbf{q}\mathbf{k}} = \int \frac{d^d \mathbf{p}}{(2\pi)^d} \frac{\epsilon [\mathbf q\!\cdot\!(\mathbf{q}-\mathbf p)]^2}{(\lambda^2 |\mathbf{p}|^2 +1)^2} \, \delta_{\mathbf{q}\mathbf{k}}.
\end{equation}
By symmetry, odd terms vanish and only even terms contribute to the integral:
\begin{equation}
\mat\Sigma_{\mathbf{q}\mathbf{k}} = 
\underbrace{\int \frac{d^d \mathbf{p}}{(2\pi)^d} \frac{\epsilon |\mathbf{q}|^4
}{(\lambda^2 |\mathbf{p}|^2 +1)^2}}_{\coloneqq \epsilon \, S_0^{(1)} \, |\mathbf{q}|^4} \, \delta_{\mathbf{q}\mathbf{k}}
+ \underbrace{\int \frac{d^d \mathbf{p}}{(2\pi)^d} \frac{\epsilon (\mathbf q\!\cdot\!\mathbf p)^2 }{(\lambda^2 |\mathbf{p}|^2 +1)^2}}_{\coloneqq \epsilon \, S_2^{(1)} \, |\mathbf{q}|^2} \, \delta_{\mathbf{q}\mathbf{k}}
.
\end{equation}

In $d{=}1$ the integrals converge and give
\begin{equation}
S_0^{(1)}=\int\!\frac{dk}{2\pi}\frac{1}{(1+\lambda^2 k^2)^2}=\frac{1}{4\lambda},\qquad
S_2^{(1)}=\int\!\frac{dk}{2\pi}\frac{k^2}{(1+\lambda^2 k^2)^2}=\frac{1}{4\lambda^3}.
\end{equation}
Therefore, for $\tau\ll t\ll \tau_r$,
\begin{equation}
\partial_t \langle h_q\rangle
=\Big[-(q^2+q^4)+\Sigma_{qq}\Big]\langle h_q\rangle
\quad\Rightarrow\quad
\partial_t \log \langle h_q\rangle
\simeq \Big(\frac{\varepsilon}{4\lambda^3}-1\Big)\,q^2
+\Big(\frac{\varepsilon}{4\lambda}-1\Big)\,q^4.
\label{eq:app-dispersion}
\end{equation}
In $d\ge 2$, both integrals can be written in hyperspherical coordinates, and the same structure holds with $S_{0,2}^{(d)}$; the integrals are UV-regularized by the microscopic cutoff (or by the lattice/bending scale).

Equation~\eqref{eq:app-dispersion} shows that fluctuations \emph{renormalize} both the tension-like ($q^2$) and bending-like ($q^4$) parts. A finite band of unstable modes appears once the $q^2$ coefficient turns positive (in 1D: $\varepsilon>4\lambda^3$), while the $q^4$ term penalizes large $q$. 
This is a \emph{short-time} (second-order cumulant/Magnus) result valid for $\tau\ll t\ll \tau_r$. 
A quantitative, all-times treatment retains the non-Markovian memory (Novikov/Furutsu), used in the main text.

\subsection{XMDS2 simulation: 1D interface with fluctuating surface tension}
Simulations were carried out with the XMDS2 program \cite{xmds2}. We simulate a one-dimensional periodic interface height field $h(x,t)$ on $x\in(0,20)$ discretized with $N_x=200$ grid points and propagated up to $t=20$ using a semi-implicit (SI) pseudospectral scheme with $80000$ time steps ($\Delta t=2.5\times10^{-4}$) and $500$ stored samples (plus the initial sample). An ensemble of $50000$ independent stochastic trajectories is generated via the \texttt{mpi-multi-path} driver. The dynamics is driven by a stochastic tension field $\xi(x,t)$ with noise strength $\epsilon$ and constants $\tau=0.2$, $\lambda=1$, $\sigma=1$, $\kappa=1$. Initial conditions are $\xi(x,0)=0$ and $h(x,0)=0.01\,\eta_0(x)$ with $\eta_0$ a unit-variance Gaussian random field.

The coupled SPDEs integrated by XMDS2 are
\begin{align}
\partial_t \xi(x,t) &= -\frac{1}{\tau}\left(1-\lambda\,\partial_x^2\right)\xi(x,t)+\frac{\sqrt{2\epsilon}}{\tau}\,\eta(x,t), \\
\partial_t h(x,t) &= -\kappa\,\partial_x^4 h(x,t)+\sigma\,\partial_x^2 h(x,t)+\partial_x\!\big(\xi(x,t)\,\partial_x h(x,t)\big)-h(x,t)^3,
\end{align}
where $\eta(x,t)$ is a Wiener noise field (XMDS \texttt{kind="wiener"}). Spatial operators are evaluated in Fourier space according to: $\partial_x\leftrightarrow -ik_x$, $\partial_x^2\leftrightarrow -k_x^2$, $\partial_x^4\leftrightarrow k_x^4$, and the $\xi$-relaxation kernel corresponds to multiplication by $-(\lambda k_x^2+1)/\tau$. The noise relaxation kernel and the bending and tension terms of the $h$ field are initialized as interaction picture ("ip") operators, while the Laplacian and first derivative in the $\partial_t(\xi(x,t)\partial_xh(x,t))$ term are initialized as explicit ("ex") operators. The multiplicative coupling is implemented as $\partial_x(\xi\partial_x h)=(\partial_x\xi)(\partial_x h)+\xi\,\partial_x^2 h$ in the right-hand side of the equations of motion. From the sampled 50000 trajectories, we calculated the average height profile, $\langle h(x,t) \rangle$, and the two-point, same-time correlation function $C(x,x';t)= \langle h(x,t)h(x',t) \rangle$. The structure factor $S(q)$ was obtained from the Fourier transform of $C(x,x',t)$, and averaged over the last 100 saved time points of the dynamics. The dominant wave vector $q^*$ was extracted from the maximum of the interpolated $S(q)$ curves on a 100 times finer grid for $q$. The interpolation was done with the ``interpolate'' function of Julia's Interpolations.jl, with the option ``BSpline(Cubic(Line(OnGrid())))''.

\subsection*{Derivation of average dynamics with Novikov--Furutsu theorem}

Here, we derive the equation Eq.~\eqref{main-eq:effective_mean_field} in the main text from the Novikov--Furutsu theorem applied to the product between the colored noise and a one-dimensional height field in d dimensions. This is done to first order in the noise intensity, corresponding to a separation of timescales between noise and the deterministic dynamics.

In Fourier space the (dimensionless) equation of motion, Eq.~\eqref{main-eq:phys_eom}, reads
\begin{equation}
\partial_t h_{\mathbf k}(t)
= -\big(k^2+k^4\big)\,h_{\mathbf k}(t)
-\int\frac{d^d \mathbf q}{(2\pi)^d}(\mathbf q\!\cdot\!\mathbf k)\,
\sigma_{\mathbf k-\mathbf q}(t)\,h_{\mathbf q}(t),
\label{eq:NF-eom-Fourier}
\end{equation}
where $k:=|\mathbf k|$. Averaging over the Gaussian noise gives
\begin{equation}
\partial_t \langle h_{\mathbf k}(t)\rangle
= -\big(k^2+k^4\big)\,\langle h_{\mathbf k}(t)\rangle
-\int\frac{d^d \mathbf q}{(2\pi)^d}(\mathbf q\!\cdot\!\mathbf k)\,
\langle \sigma_{\mathbf k-\mathbf q}(t)\,h_{\mathbf q}(t)\rangle.
\label{eq:NF-avg-start}
\end{equation}

For a zero-mean Gaussian field $\sigma$ with correlator
$\Gamma_{\mathbf p,\mathbf p'}(t-s):=\langle\sigma_{\mathbf p}(t)\sigma_{\mathbf p'}(s)\rangle$,
the Novikov--Furutsu theorem \cite{novikov,furutsu} yields
\begin{align}
\big\langle \sigma_{\mathbf k-\mathbf q}(t)\,h_{\mathbf q}(t)\big\rangle
&=\int\frac{d^d \mathbf k'}{(2\pi)^d}\int_{t_0}^{t}\!ds\;
\big\langle \sigma_{\mathbf k-\mathbf q}(t)\,\sigma_{\mathbf k'}(s)\big\rangle
\Big\langle \frac{\delta h_{\mathbf q}(t)}{\delta\sigma_{\mathbf k'}(s)}\Big\rangle
\nonumber\\
&=\int_{t_0}^{t}\!ds\;
\Gamma_{\mathbf k-\mathbf q,\,-\mathbf k+\mathbf q}(t-s)\,
\Big\langle \frac{\delta h_{\mathbf q}(t)}{\delta\sigma_{-\mathbf k+\mathbf q}(s)}\Big\rangle,
\label{eq:NF-Novikov}
\end{align}
where in the last step we used the Fourier-space structure
$\Gamma_{\mathbf p,\mathbf p'}\propto \delta_{\mathbf p,-\mathbf p'}$.

To evaluate the response functional
$\delta h_{\mathbf q}(t)/\delta\sigma_{-\mathbf k+\mathbf q}(s)$,
one may differentiate the time-ordered formal solution
(Eq.~\eqref{eq:app-formal}) with respect to $\sigma$.
The chain rule acts on the time-ordered exponential because $\sigma$ enters
$\mat\Lambda(t')$ under the time ordering. This gives
\begin{equation}
    \Biggl\langle \frac{\delta h_{\mathbf q}(t)}{\delta\sigma_{-\mathbf k+\mathbf q}(s)} \Biggr\rangle = \Biggl\langle \int\frac{d^d \mathbf n}{(2\pi)^d}\Big\lceil\exp\!\Big\{\int_{s}^{t}\!\!dt'\,[\mat J+\mat\Lambda(t')]\Big\}\Big\rceil_{\mathbf q,~\mathbf n+\mathbf q-\mathbf k}(\mathbf k-\mathbf n-\mathbf q)\cdot\mathbf n~ h_{\mathbf n}(s) \Biggr\rangle.
\end{equation}

We assume that at first order in the noise intensity, the dynamics from $t'$ follow the deterministic motion with $\mat\Lambda(s)=0$.
This corresponds to a separation of time scales, for which the noise decorrelates on time $\tau$ while the deterministic relaxation time
$\tau_r$ is larger, $\tau_r > \tau$. We then approximate the propagation
from $s$ to $t$ by the deterministic evolution generated by $\mat J$.
With $J_{\mathbf q\mathbf q}=-(|\mathbf q|^2+| \mathbf q |^4)$ this gives
\begin{equation}
\Big\langle \frac{\delta h_{\mathbf q}(t)}{\delta\sigma_{-\mathbf k+\mathbf q}(s)}\Big\rangle
\simeq
-\big(\mathbf q\!\cdot\!\mathbf k\big)\,
e^{J_{\mathbf q\mathbf q}(t-s)}\,
\langle h_{\mathbf k}(s)\rangle.
\label{eq:NF-response}
\end{equation}

Substituting \eqref{eq:NF-response} into \eqref{eq:NF-Novikov} yields
\begin{equation}
\big\langle \sigma_{\mathbf k-\mathbf q}(t)\,h_{\mathbf q}(t)\big\rangle
\simeq
-\int_{t_0}^{t}\!ds\;
\Gamma_{\mathbf k-\mathbf q,\,-\mathbf k+\mathbf q}(t-s)\,
\big(\mathbf q\!\cdot\!\mathbf k\big)\,
e^{J_{\mathbf q\mathbf q}(t-s)}\,
\langle h_{\mathbf k}(s)\rangle.
\label{eq:NF-sigh}
\end{equation}

Inserting \eqref{eq:NF-sigh} into \eqref{eq:NF-avg-start}, we obtain the
(non-Markovian) integro-differential equation
\begin{equation}
\partial_t \langle h_{\mathbf k}(t)\rangle
= -\big(k^2+k^4\big)\,\langle h_{\mathbf k}(t)\rangle
+\int\frac{d^d \mathbf q}{(2\pi)^d}\big(\mathbf q \cdot \mathbf k\big)^2
\int_{t_0}^{t}\!ds\;
\Gamma_{\mathbf k-\mathbf q,\,-\mathbf k+\mathbf q}(t-s)\,
e^{J_{\mathbf q\mathbf q}(t-s)}\,
\langle h_{\mathbf k}(s)\rangle.
\label{eq:NF-mean-eq}
\end{equation}
This is the Novikov--Furutsu result for the average mode dynamics. In the
white-noise (small-$\tau$) limit, the memory kernel becomes short-ranged and
\eqref{eq:NF-mean-eq} reduces to the Markovian generator renormalization found
from the short-time cumulant/Magnus analysis in the previous section.

\subsection*{Derivation of dispersion relation for average of height profile in Fourier space}

In this section, we derive the dispersion relation $\gamma(q)$ for the dynamics of the average height profile in one dimension given by Eq.~\eqref{main-eq:effective_mean_field}, used to plot the quantities in Fig.~\ref{main-fig:disp}. We start by expressing the equation Eq.~\eqref{main-eq:effective_mean_field} in Laplace form in time. The resulting equation for the Laplace mode $ \langle \tilde{h}_q(s) \rangle $ reads
\begin{equation}
    s\langle \tilde{h}_q(s) \rangle -\langle h_q(0) \rangle = J_q\langle \tilde{h}_q(s) \rangle + \frac{\epsilon q^2}{\tau}\tilde{K}_q(s)\langle \tilde{h}_q(s) \rangle,
\end{equation}
where
\begin{equation}
    \tilde{K}_q(s) = \int_{-\infty}^{+\infty} \frac{dn}{2\pi} \frac{(n-q)^2}{\left(\lambda ^2 n^2+1\right) \left((n-q)^2+(n-q)^4+\frac{\lambda ^2 n^2+1}{\tau }+s\right)}.
\end{equation}

Expanding $\tilde{K}_q(s)$ in first order in $s$ gives, after shifting $n$ by $q$:
\begin{equation}
\begin{aligned}
    \tilde{K}_q(s)=&\int_{-\infty}^{+\infty} \frac{dn}{2\pi}\frac{n^2}{\left(\lambda ^2 (q+n)^2+1\right) \left(\frac{\lambda ^2 (q+n)^2+1}{\tau }+n^4+n^2\right)}\\
    &-\int_{-\infty}^{+\infty} \frac{dn}{2\pi}\frac{n^2 s}{\left(\lambda ^2 (q+n)^2+1\right) \left(\frac{\lambda ^2 (q+n)^2+1}{\tau }+n^4+n^2\right)^2}.
\end{aligned}
\end{equation}

For brevity, we name $A_0(\lambda,\tau,q)$ and $A_0(\lambda,\tau,q)$ respectively the zeroth- and first-order term of the s expansion above.

We thus have
\begin{equation}
\begin{aligned}
    \langle \tilde{h}_q(s) \rangle & = \frac{\langle h_q(0) \rangle }{s-J_q-\frac{\epsilon q^2}{\tau}A_0(\lambda,\tau,q)-s\frac{\epsilon q^2}{\tau}A_1(\lambda,\tau,q)} \\
    & = \left( 1-\frac{\epsilon q^2}{\tau}A_1(\lambda,\tau,q) \right) \frac{\langle h_q(0) \rangle}{s+\frac{-J_q-\frac{\epsilon q^2}{\tau}A_0(\lambda,\tau,q)}{1-\frac{\epsilon q^2}{\tau}A_1(\lambda,\tau,q)}}.
\end{aligned}
\end{equation}

From this, we read the dispersion relation as
\begin{equation}
    \gamma(q) = \frac{J_q+\frac{\epsilon q^2}{\tau}A_0(\lambda,\tau,q)}{1-\frac{\epsilon q^2}{\tau}A_1(\lambda,\tau,q)}.
    \label{eq:disp-relation-laplace}
\end{equation}

To evaluate the dispersion relation in Eq.~\eqref{eq:disp-relation-laplace}, we numerically evaluated the integrals in $A_0(\lambda,\tau,q)$ and $A_1(\lambda,\tau,q)$ with the Python Scipy library. 
The dispersion relation ($\gamma(q)$) was obtained for $\lambda = 1.0,~\tau=0.2$ over a grid of 2000 evenly spaced points in the interval $k\in[0,1.1]$. The heatmap for the marginal wave number as a function of $\lambda$ and $\tau$ was produced over a grid of 100 points, both for $\lambda \in [0,1]$ and $\tau\in[0,1]$; for every couple $(\tau,~\lambda)$, the marginal wave number was obtained from the dispersion relation sampled over 200 evenly spaced points in the interval $k\in[0,20]$.

\bibliography{bibl}